# PKC-PC: A Variant of the McEliece Public Key Cryptosystem based on Polar Codes

Reza Hooshmand, Masoumeh Koochak Shooshtari, and Mohammad Reza Aref

*Abstract*— **Polar codes are novel and efficient error correcting codes with low encoding and decoding complexities. These codes have a channel dependent generator matrix which is determined by the code dimension, code length and transmission channel parameters. This paper studies a variant of the McEliece public key cryptosystem based on polar codes, called "PKC-PC". Due to the fact that the structure of polar codes' generator matrix depends on the parameters of channel, we used an efficient approach to conceal their generator matrix. Then, by the help of the characteristics of polar codes and also introducing an efficient approach, we reduced the public and private key sizes of the PKC-PC and increased its information rate compared to the McEliece cryptosystem. It was shown that polar codes are able to yield an increased security level against conventional attacks and possible vulnerabilities on the code-based public key cryptosystems. Moreover, it is indicated that the security of the PKC-PC is reduced to solve NP-complete problems. Compared to other post-quantum public key schemes, we believe that the PKC-PC is a promising candidate for NIST post-quantum crypto standardization.**

## I. Introduction and Motivation

It has been revealed that the conventional used public key cryptosystems, whose security are based on the difficulty of discrete logarithm or factoring problems, are broken by the quantum computers in polynomial-time [1]. One of the important categories of cryptosystems which can resist quantum computer-based attacks is code-based cryptosystems. These kinds of cryptosystems can be considered as alternatives to the conventional public key cryptosystems, such as RSA and ElGamal [2]. The security of most of these cryptosystems relies on the hardness of some conventional problems in coding theory [3]. For example, it was previously shown that the decoding of a linear code with no clear structure is NP-complete problem [4]. The first public key code-based cryptosystem originally proposed based on the binary Goppa codes is McEliece cryptosystem [5]. This cryptosystem applies a binary Goppa codes' generator matrix, a scrambling matrix, and a permuting matrix as the private key. The scrambling and permuting matrices are employed to convert the private key into the public matrix. The McEliece cryptosystem applies the generator matrix and encodes the information vector into the public code's codewords. Compared to the conventional public key cryptosystems, McEliece cryptosystem has low complexity encryption/ decryption algorithms. Nevertheless, due to the use of binary Goppa codes, this cryptosystem has two major weaknesses [6]: (i) low transmission rate, and (ii) huge key size.

One of efficient approaches to resolve the weaknesses of McEliece cryptosystem is exchanging binary Goppa codes with the other linear block codes. However, such replacement can yield serious flaws in its security level. Thus far, several schemes have been proposed to dominate the weaknesses of McEliece scheme by exchanging the Goppa codes with the different linear codes such as generalized Reed–Solomon (GRS) codes [7], Reed–Muller codes [8], quasi cyclic low density parity check (QC-LDPC) codes [9-11], wild Goppa codes [12, 13], p-adic Goppa codes [11, 14], moderate density parity check (MDPC) codes [15, 16], convolutional codes [17] and more recently low density lattice codes (LDLCs) [18]. Some of these suggested yields decrease the public key length while keeping the same security level against the conventional attacks. However, most of them exposed the McEliece cryptosystem to security threats and yield serious flaws in its security level. For example, public key schemes based on GRS and Reed-Muller codes were broken in [19] and [20], respectively. A number of versions of LDPC code-based schemes [9, 10] have been successfully cryptanalyzed with efficient attacks in [21, 22]. Some of the parameters that can be found in public key schemes based on wild Goppa codes [12, 13] have been successfully cryptanalyzed in [23, 24]. In addition, the convolutional code-based scheme [17] was successfully cryptanalyzed by Landais and Tillich in [25]. Moreover, the cryptosystems based on p-adic Goppa codes [11, 14] were broken in [26].

Polar codes [27] are novel family of codes which, by the help of successive cancellation (SC) decoding, can attain the information theoretic bounds in channel coding. Up to now, by applying the properties of polar codes, many attempts have been made to achieve secrecy in information theoretic security [28]. However, several researches have been carried out in recent years to introduce the polar code-based cryptographic

The material in this paper was presented in part at the Eleventh International ISC Conference on Information Security and Cryptology (ISCISC 2014), Tehran, Iran, September 2014. This work was supported in part by Iranian National Science Foundation (INSF) under Grant 92.32575.

Reza Hooshmand is with the Department of Electrical Engineering, Shahid Sattari Aeronautical University of Science and Technology, Tehran 1384663113, Iran (e-mail: rhooshmand@ssau.ac.ir).

Masoumeh Koochak Shooshtari is with the Faculty of Electrical Engineering, K. N. Toosi University of Technology, Tehran 16315-1355, Iran (e-mail: m-koochak@ee.kntu.ac.ir).

Mohammad Reza Aref is with the Department of Electrical Engineering, Sharif University of Technology, Tehran 11365/8639, Iran, (e-mail: aref@sharif.edu)

schemes. The employing of polar codes in the construction of symmetric key cryptosystems is presented in [29, 30]. Also, by performing the properties of finite-length polar codes, physical layer encryption (PLE) schemes are presented in [31, 32] to make secure communication between legitimate partners against active and passive adversaries. In [33, 34], polar code-based versions of McEliece cryptosystem are introduced. However, in [35], Bardet et.al have presented a key-recovery attack against recently Shrestha-Kim [34] polar code-based public key scheme. This type of key-recovery attack makes Shrestha-Kim scheme possible to obtain all the needed information for decryption of any message. In addition, it allows breaking such version of McEliece cryptosystem for the parameters suggested by Shrestha and Kim. Although, the only approach to prevent this type of attack is to find polar code parameters for which finding minimum weight codewords is impossible. This requires changing significantly the parameters proposed in [34]. Moreover, it is shown that such key recovery attack is not directly applicable to another public key scheme presented in [33] which is based on random subcode of the polar code [35].

In this paper, a secure and reliable polar code-based public key scheme is introduced to resolve the weaknesses of the original McEliece scheme. Due to the following reasons, it is rational to study whether polar codes are convenient to use in a McEliece-like public key scheme: (i) polar codes have better error correction capability than Goppa code which allows decreasing of the key length of the polar code-based scheme; (ii) the SC decoding of polar codes is performed faster than decoding of Goppa codes which yields decreasing the computational complexity of the decryption process; (iii) the polar codes have large equivalent family by which attacking the polar code-based scheme to dissolve the code equivalence problem is infeasible; and (iv) since polar codes have large hull and large permutation group, executing support splitting algorithm (SPA) algorithm is doomed to fail for polar code-based scheme [35]. In the PKC-PC, we conceal the polar codes' generator matrix by random selection of its $k$ rows from an $n \times n$ square matrix. The advantage of random selection in this method is that the opponent cannot obtain the needed data to decode the predetermined polar codeword. In addition, by exploiting the properties of polar codes, we use the encryption matrix of the PKC-PC in systematic form. Moreover, the nonsingular matrix is obtained from the generator matrix of employed polar code. These proceedings yield a reduction in the private and public key sizes and also have an increment in the security level.

The rest of this paper is organized as follows. Section II describes the characteristics of polar codes. We discuss the idea of applying polar codes in the construction of the PKC-PC in Section III. Moreover, the design issues of our scheme are explained. The efficiency level of the PKC-PC is assessed in Section IV. In fact, we compute the error performance, key size and the computational complexity of the PKC-PC and then compare it with original McEliece and other McEliece-like schemes. To investigate the security level, we consider the security reduction in Section V. Also, we show that the PKC-PC has high enough security level by choosing the proper values of the parameters. Finally, the conclusion of this paper is presented in Section VI.

## II. Polar Codes

Polar codes are very powerful category of linear codes that demonstrably attain any Binary-input Discrete Memoryless Channel's (B-DMC) capacity, e.g., Binary Erasure Channel (BEC) [27] and Binary Symmetric Channel (BSC). Let $W : \mathcal{X} \to \mathcal{Y}$ be a B-DMC. Consider $\mathcal{X} = \{0, 1\}$ as an input alphabet, $\mathcal{Y}$ as an output alphabet and $\{W(\boldsymbol{y}|\boldsymbol{x}), \boldsymbol{x} \in \mathcal{X}, \boldsymbol{y} \in \mathcal{Y}\}$ as the transition probabilities of $W$. Let us consider $I(W) \triangleq \sum_{y\in\mathcal{Y}} \sum_{\boldsymbol{x}\in\mathcal{X}} \frac{1}{2} W(\boldsymbol{y}|\boldsymbol{x}) \log \frac{W(y|x)}{\frac{1}{2}W(y|0)+\frac{1}{2}W(y|1)}$ and $Z(W) \triangleq \sum_{y\in\mathcal{Y}} \sqrt{W(\boldsymbol{y}|0)W(\boldsymbol{y}|1)}$ for $W$, where $I(W) \in [0, 1]$ is called the capacity for $W$ and hence performed for measuring the rate. Also, $Z(W) \in [0, 1]$ is called the Bhattacharyya parameter of $W$ and applied to measure the reliability. In this case, $I(W) \approx 1$ iff $Z(W) \approx 0$, also $I(W) \approx 0$ iff $Z(W) \approx 1$. If $W$ is a BEC with erasure probability $\epsilon$, i.e., BEC($\epsilon$), then we have $Z(W) = \epsilon$ and $I(W) = 1 - Z(W) = 1 - \epsilon$. Let $\left\{W_n^{(i)}: 1 \le i \le n\right\}$ be a set of polarized channels, called sub-channels or bit-channels, with indices '$i$' that are obtained by applying the channel polarization process on the $n$ independent copies of a B-DMC $W$. If $n$ is large enough, the $n$ sub-channel's capacities $\left\{I\left(W_n^{(i)}\right), 1 \le i \le n\right\}$ and the $n$ sub-channel's Bhattacharya parameters $\left\{Z\left(W_n^{(i)}\right), 1 \le i \le n\right\}$ will be 0 or 1. Let $\mathcal{I}_n = \{i, 1 \le i \le n\}$ be an $n$ sub-channel indices set. Consider $\mathcal{A} \subset \mathcal{I}_n$ as a $k$-element information set and $\mathcal{A}^c \subset \mathcal{I}_n$ as an $(n-k)$-element frozen (fixed) set. For all $i \in \mathcal{A}$, $j \in \mathcal{A}^c$, we have $Z\left(W_n^{(i)}\right) \le Z\left(W_n^{(j)}\right)$ and $I\left(W_n^{(i)}\right) \ge I\left(W_n^{(j)}\right)$. In fact, for $n$ sub-channels, $nI(W)$ sub-channels (with indices in $\mathcal{A}$) become noiseless or reliable and $n(1 - I(W))$ sub-channels (with indices in $\mathcal{A}^c$) become unreliable or noisy [27].

### A. Constructing the Generator and Parity-Check Matrices

Consider $n = 2^m$, $m \ge 1$ and $G_2 = \begin{bmatrix} 1 & 0 \\ 1 & 1 \end{bmatrix}$. Also, consider the $m$-th Kronecker product $G_2^{\otimes m}$ which yields an $n \times n$ matrix. One interesting property of matrix $G_2^{\otimes m}$ for polar codes is shown in Remark 1 [36]:

*Remark 1:* Let $(G_n)_{\mathcal{A}\mathcal{A}}$ denotes the submatrix of $G_n$ consisting of the array of elements $G_{i,j}$, $i, j \in \mathcal{A}$. Any submatrix $(G_n)_{\mathcal{A}\mathcal{A}}$, $\mathcal{A} \subset \{1, \cdots, n\}$ is also a lower-triangular matrix and has 1s on the diagonal, so it is also nonsingular (invertible).

Given the rate $R < I(W)$ and dimension $k = nR$, a $k \times n$ generator matrix $G_{\mathcal{A}}$ is obtained for any polar code of length $n$ and dimension $k$ with the subsequent steps [37]:

1) First, the rows of $G_n$ are labeled from the first to the last row as $i = 1, 2, \cdots, n$. For BEC($\epsilon$), $Z\left(W_n^{(i)}\right)$, $1 \le i \le n$ are obtained as follows: (i) $\forall 1 \le i \le l$, $l = 1, 2, 2^2, \cdots, 2^{m-1}$ $Z\left(W_{2l}^{(i)}\right) = 2Z\left(W_l^{(i)}\right) - Z^2\left(W_l^{(i)}\right)$; (ii) $\forall k + 1 \le i \le 2l$, $Z\left(W_{2l}^{(i)}\right) = Z^2\left(W_l^{(i-l)}\right)$. A permutation $\pi_n = (i_1, \ldots, i_n)$ is formed for $n$ sub-channel indices set $\mathcal{I}_n = \{1, 2, \cdots, n\}$ such that $Z\left(W_n^{(i_j)}\right) \leq Z\left(W_n^{(i_k)}\right)$, $1 \leq j < k \leq n$.

2) The information set $\mathcal{A} \subset \mathcal{I}_n$ is obtained whose indices of sub-channels correspond to $k$ leftmost indices in $\pi_n$, i.e., $i_1, \ldots, i_k$. The $k \times n$ generator matrix $G_{\mathcal{A}}$ is obtained by choosing $k$ rows of $G_n$ related to the information set indices $\mathcal{A}$.
3) The frozen set $\mathcal{A}^c \subset \mathcal{I}_n$ is obtained whose indices of sub-channels correspond to $(n-k)$ rightmost indices in $\pi_n$, i.e. $i_{k+1}, i_{k+2}, \ldots, i_n$. The $(n-k) \times n$ frozen matrix $G_{\mathcal{A}^c}$ is generated by choosing $(n-k)$ rows of $G_n$ related to the frozen indices set $\mathcal{A}^c$.

In the $(n,k)$ non-systematic polar codes, an input vector $\boldsymbol{u} = (u_1, u_2, \cdots, u_n) = (\boldsymbol{u}_{\mathcal{A}}, \boldsymbol{u}_{\mathcal{A}^c})$ consists of $k$-bit information subvector $\boldsymbol{u}_{\mathcal{A}} = (u_i, i \in \mathcal{A})$ and $(n-k)$-bit frozen (fixed) subvector $\boldsymbol{u}_{\mathcal{A}^c} = (u_i, i \in \mathcal{A}^c)$. The information subvector $\boldsymbol{u}_{\mathcal{A}}$ consists of information data that can be changed in transmission process. The frozen subvector consists of fixed values known to decoder [36]. Polar codes are defined in terms of an invertible map $G_n$ via $\boldsymbol{x} = \boldsymbol{u}G_n$. The matrix $G_n = B_n G_2^{\otimes m}$, where $B_n$ is a bit-reversal permutation matrix defined in [27]. The construction of polar codes' parity check matrix is characterized as the lemma 1 [38]:

*Lemma 1:* Let $\mathcal{A}$ be an information set and let $\mathcal{A}^c$ be a frozen set of an $(n,k)$ polar code. Let $G_n = B_n G_2^{\otimes m}$ be an $n \times n$ matrix consist of the generator matrix $G_{\mathcal{A}}$ and the frozen matrix $G_{\mathcal{A}^c}$. Also, assuming that frozen vector $\boldsymbol{u}_{\mathcal{A}^c}$ is all-zero vector. In this case, the parity check matrix $H_{n \times r}$, $r = n - k$ is constructed by selecting the columns of $G_n$ with indices in $\mathcal{A}^c$.

*Proof.* Similar to the Lemma 1's proof in [38].

### B. Polar Encoding

In the polar encoding process, the input vector $\boldsymbol{u} = (\boldsymbol{u}_{\mathcal{A}}, \boldsymbol{u}_{\mathcal{A}^c})$ is converted to $n$-bit codeword $\boldsymbol{x} = \boldsymbol{u}_{\mathcal{A}} G_{\mathcal{A}} + \boldsymbol{u}_{\mathcal{A}^c} G_{\mathcal{A}^c} = \boldsymbol{u}_{\mathcal{A}} G_{\mathcal{A}} + c$, where $c \triangleq \boldsymbol{u}_{\mathcal{A}^c} G_{\mathcal{A}^c}$ is a fixed vector. The code rate is obtained as $R = |\boldsymbol{u}_{\mathcal{A}}|/|\boldsymbol{x}| = |\mathcal{A}|/n$. The information vector is sent across the noiseless sub-channels at a rate close to one. In addition, the frozen (fixed) vector is sent across the noisy sub-channels at a rate close to zero [27].

### C. Successive Cancelation (SC) Decoding

Consider $\boldsymbol{x}$ as an $n$-bit polar code's codeword that is sent along the $n$ sub-channels. Consider $\boldsymbol{y} = y_1^n$ as a related channel output vector. The main aim of SC decoding is to compute the evaluated input vector $\hat{\boldsymbol{u}}$ using information set $\mathcal{A}$, frozen vector $\boldsymbol{u}_{\mathcal{A}^c}$ and channel output vector $\boldsymbol{y}$. In fact, for $W_n^{(i)}, i = 1, 2, \cdots, n$, the SC decoding computes the likelihood ratio (LR) of bits of input vector $u_i$ given $\boldsymbol{y}$ and the past obtained information bits $\hat{u}_1^{i-1}$ as $L_n^{(i)} = \frac{W_n^{(i)}(y_1^n, \hat{u}_1^{i-1} | u_i = 0)}{W_n^{(i)}(y_1^n, \hat{u}_1^{i-1} | u_i = 1)}$. The input vector bits are obtained with the help of SC decoding as follows: (i) $\forall i \in \mathcal{A}^c$, $\hat{u}_i = u_i$; (ii) $\forall i \in \mathcal{A}$, $\hat{u}_i = h_i(y_1^n, \hat{u}_1^{i-1})$. The decision functions $h_i: \mathcal{Y}^n \times \mathcal{X}^{i-1} \to \mathcal{X}, i \in \mathcal{A}$ are obtained for all $y_1^n \in \mathcal{Y}^n$, $\hat{u}_1^{i-1} \in \mathcal{X}^{i-1}$ as follows: (i) $\forall\, L_n^{(i)} \geq 1$, $h_i(y_1^n, \hat{u}_1^{i-1}) = 0$ and (ii) otherwise, $h_i(y_1^n, \hat{u}_1^{i-1}) = 1$. The upper bound on error probability with the SC decoder is obtained as $P_e \leq \sum_{i \in \mathcal{A}} Z\left(W_n^{(i)}\right)$ for any B-DMC $W$ [27]. Moreover, it is shown that reliable communication can be achieved under SC decoding by satisfying the inequality (1) [39],

$$R < I(W) - n^{-1/\mu}, \tag{1}$$

where $\mu$ is named scaling exponent and its values depend on the channel type. For example, we have $\mu \approx 3.627$ for BEC. The maximum possible $R$ satisfying (1) is called *cutoff rate* and shown by $R_0$.

## III. The Proposed Polar Code-based Public Key Scheme

Here, first efficient techniques are presented to categorize the sub-channels and also conceal the polar codes' generator matrix. Then, we explain the construction of the PKC-PC.

### A. Good and Bad Sub-Channels

For the PKC-PC, we categorize all $n$ sub-channels $\{W_n^{(i)}: 1 \leq i \leq n\}$ into good and bad sub-channels as definitions 1 and 2 [29]:

*Definition 1.* The $nR_0$ sub-channels are considered as *good sub-channels* if they have minimum Bhattacharya parameters among all $n$ sub-channels, i.e., minimum error probability. The good sub-channels' indices are related to the $nR_0$ leftmost indices of $\pi_n$ and shown as $\mathcal{G}_n(W, R_0) = \left\{i \in \mathcal{I}_n : \pi_n(i) \in \{i_1, i_2, \cdots, i_{nR_0}\}\right\}$. ■

*Definition 2.* The $n(1-R_0)$ sub-channels are considered as *bad sub-channels* if they have maximum Bhattacharya parameters among all $n$ sub-channels, i.e., maximum error probability. The bad sub-channels' indices are related to the $n(1-R_0)$ rightmost indices of $\pi_n$ and shown as $\mathcal{B}_n(W, R_0) = \left\{i \in \mathcal{I}_n : \pi_n(i) \in \{i_{nR_0+1}, i_{nR_0+2}, \cdots, i_n\}\right\}$. ■

In the PKC-PC, we consider the transmission over the noiseless insecure channel. In this case, all $n$ sub-channels are considered as the good sub-channels. Therefore, we can use high transmission rate, e.g. 0.9, in the PKC-PC. In fact, the information rate of PKC-PC is increased significantly compared to the McEliece cryptosystem.

### B. Concealing the Generator Matrix

To hide the polar codes' generator matrix, an efficient approach is being proposed in the following steps by which an adversary cannot obtain the concealed polar codes' generator matrix:

1) First, $k$ indices are chosen randomly from $\mathcal{G}_n(W, R_0)$. In fact, this process is related to the random selection of $k$ sub-channels from the set of good sub-channels. The arbitrarily $k$ chosen indices from $\mathcal{G}_n(W, R_0)$ are named as the *secret information set* and shown by $\mathcal{A}(s)$. The *secret generator matrix*, $G_{\mathcal{A}(s)}$, is constructed as a $k \times n$ submatrix of $G_n$ with $k$ selected rows corresponding to $\mathcal{A}(s)$.
2) The *secret frozen set*, $\mathcal{A}^c(s)$, is a subset of $\mathcal{I}_n$ whose $(n-k)$ indices are not selected from $\mathcal{I}_n$ in step 1. Also, the *secret frozen matrix* $G_{\mathcal{A}^c(s)}$ is constructed as an $(n-k) \times n$ submatrix of $G_n$ whose rows are selected related to $\mathcal{A}^c(s)$.

In this way, the secret generator matrix $G_{\mathcal{A}(s)}$ cannot be recovered by the adversary even if $\epsilon$, $n$ and $k$ are known. In fact, by concealing the $G_{\mathcal{A}(s)}$, the attacker cannot recover the estimated input vector $\hat{\boldsymbol{u}}$ from the channel output vector $\boldsymbol{y}$ in polynomial-time. Fig. 1 shows the proposed concept of concealing the generator matrix and encoding to enhance the security based on an $(n, k)$ polar code.

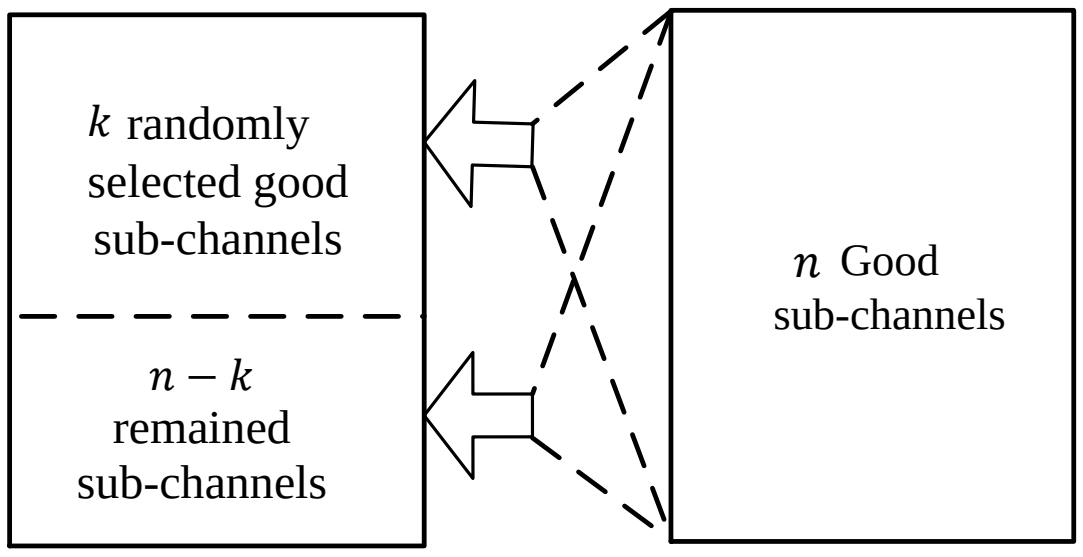

(a)

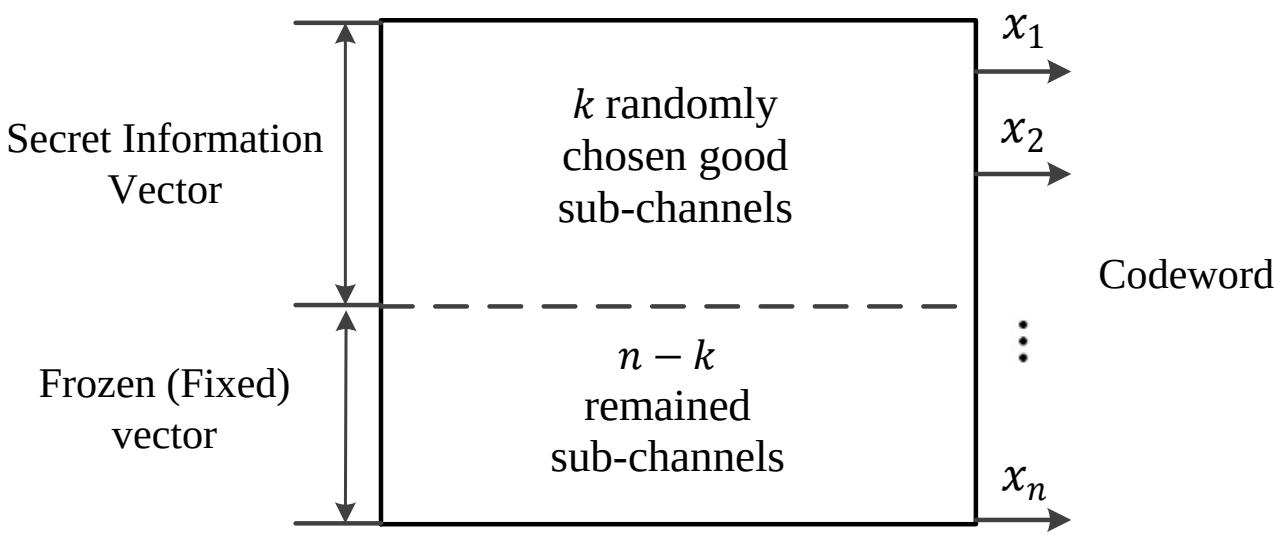

(b)

Fig. 1. The idea of providing security and encoding based on an $(n, k)$ polar code. (a) The $k$ sub-channels are randomly chosen from $n$ good sub-channels. (b) The secret information vector is sent through $k$ randomly chosen good sub-channels. Moreover, the fixed bits (zeros) are sent across $(n-k)$ non-selected sub-channels.

As observed in Fig. 1(a), $k$ sub-channels are arbitrarily chosen from the good sub-channels to conceal the generator matrix $G_{\mathcal{A}(s)}$. The idea in Figure 1(b) is to transmit the secret information vector, denoted by $\boldsymbol{u}_{\mathcal{A}(s)}$, across $k$ randomly chosen good sub-channels while transmitting the fixed vector, denoted by $\boldsymbol{u}_{\mathcal{A}^c(s)}$, through the $(n-k)$ remained sub-channels. Since the error performance of polar codes does not depend on the case in which $\boldsymbol{u}_{\mathcal{A}^c(s)}$ is chosen, it makes no difference how this vector is selected. Hence, an $(n-k)$-bit all-zero vector is considered as $\boldsymbol{u}_{\mathcal{A}^c(s)}$ in the encrypting/decrypting algorithms of the PKC-PC to make its simplified structure.

## C. Key Generation

The key generating algorithm performs as follows:

1) A secret generator matrix $G_{\mathcal{A}(s)}$ is generated (see Sec. III.B).
2) A $k \times k$ scrambling matrix $S$ is generated by extracting a submatrix $(G_n)_{\mathcal{A}(s)\mathcal{A}(s)}$ from $G_n$ (see Sec. III.B).
3) An $n \times n$ binary permuting matrix $P = [P'|P'']$ is generated. In this case, $P'$ is the $n \times k$ submatrix in which $k$ '1's are respectively placed, one in each of $j$-th, $j \in \mathcal{A}(s)$ rows of its $k$ columns. In addition, $P''$ is the $n \times (n-k)$ submatrix whose $(n-k)$ '1's are randomly placed in its $(n-k)$ columns such that by evaluating the positions of 1s in $P'_{n\times k}$, the permuting property of $P_{n\times n}$ is preserved.
4) The encryption matrix is constructed as $G' = S^{-1} G_{\mathcal{A}(s)} P$.

## D. Private Key

The set of private key is $\mathcal{K}_{sec} = \{\mathcal{A}^c(s), P\}$. In the PKC-PC, given that the construction of $G_{\mathcal{A}(s)}$ and $S$ are based on $\mathcal{A}(s)$, it is possible to save $\mathcal{A}(s)$ instead of $G_{\mathcal{A}(s)}$ and $S$. Also, the set $\mathcal{A}^c(s)$ is complement to $\mathcal{A}(s)$ and needs less memory to store, hence it is enough to save $\mathcal{A}^c(s)$ instead of $\mathcal{A}(s)$. This concept dramatically leads to reduction of the private key length (see Sec. IV.B). Another element of $\mathcal{K}_{sec}$ is the permutation matrix $P_{n\times n}$ whose construction is defined in Sec. III.C.

## E. Public Key

The public key is obtained as $\mathcal{K}_{pub} = G' = S^{-1} G_{\mathcal{A}(s)} P = S^{-1} G''$, where $G'' = G_{\mathcal{A}(s)} P$ is a $k \times n$ matrix. Each of $k$ leftmost columns of $G_{\mathcal{A}(s)}$ are ordered related to $j \in \mathcal{A}(s)$ indices by multiplying $G_{\mathcal{A}(s)}$ and permuting matrix $P$ together. In such way, $G'' = [S|G''']$ includes two submatrices: a $k \times k$ nonsingular submatrix $S = (G_n)_{\mathcal{A}(s)\mathcal{A}(s)}$ and a $k \times (n-k)$ submatrix $G'''$. Hence, the public key is obtained as a $k \times n$ matrix $\mathcal{K}_{pub} = S^{-1} G'' = [I_k|Q]$, where $I_k$ is a $k \times k$ identity submatrix and $Q = S^{-1} G'''$ is a $k \times (n-k)$ submatrix. With the help of this method, the required memory to save the public key $\mathcal{K}_{pub}$ is $k(n-k)$ bits instead of $kn$ bits. In fact, it suffices to store $k \times (n-k)$ matrix $Q$ instead of $\mathcal{K}_{pub}$. In this way, the large key length problem of the McEliece cryptosystem can be solved. It should be noted that in this case, the memory requirement of $\mathcal{K}_{pub}$ is further reduced by increasing the information rate. Also, by employing the CCA2-secure Kobara-Imai's $\gamma$-conversion [40] for the PKC-PC, the systematic encryption matrix $G'$ does not decrease its security level against adaptive chosen ciphertext attacks.

## F. Encryption

Bob first randomly selects a code in the family of equivalent $(n, k)$ polar codes by randomly choosing $k$ indices from the good sub-channel indices. Then, he considers the indices of $k$ selected good sub-channels as $\mathcal{A}(s)$ and constructs $G_{\mathcal{A}(s)}$ for the selected polar code. Also, he constructs two other secret matrices; a $k \times k$ scrambling matrix $S$ and an $n \times n$ permuting matrix $P$ as in the aforementioned processes in Sec. III.C. In addition, Bob generates a public key as $G' = S^{-1} G_{\mathcal{A}(s)} P$. Besides, Alice obtains $G'$ from the public directory and separates the message into $k$-bit blocks $\boldsymbol{m}$. At last, Alice performs the encryption algorithm as $\boldsymbol{c} = \boldsymbol{m}G' + \boldsymbol{e}$, where $\boldsymbol{e}$ is an arbitrary error vector such that $w_H(\boldsymbol{e}) < t$.

## G. Decryption

The ciphertext $\boldsymbol{c}$ is decrypted according to the following steps:

1) First, $\boldsymbol{c}' = \boldsymbol{c}P^{-1} = \boldsymbol{m}S^{-1} G_{\mathcal{A}(s)} + \boldsymbol{e}P^{-1}$ is computed, where $P^{-1}$ is the inverse of the permutation matrix $P$. Given that $P$ is a permutation matrix, we have $w_H(\boldsymbol{e}P^{-1}) = w_H(\boldsymbol{e})$.

Therefore, $\boldsymbol{c}' = c'^n_1$ is a codeword in the polar code previously chosen and Bob can correct the intentional errors with the help of SC decoding to recover $\boldsymbol{m}S^{-1}$. Since $\boldsymbol{u}_{\mathcal{A}^c(s)}$ is full-zero vector, the set $\{\mathcal{A}(s), \boldsymbol{c}'\}$ is noticed as the SC decoder's input (See Fig. 2).

2) The input vector of encoder, $\boldsymbol{u} = (\boldsymbol{u}_{\mathcal{A}(s)}, \boldsymbol{u}_{\mathcal{A}^c(s)}) = (\boldsymbol{m}S^{-1}, 0)$, is evaluated with the help of the SC decoding as follows: (i) $\forall i \in \mathcal{A}^c(s)$, $\hat{u}_i = 0$; (ii) $\forall i \in \mathcal{A}(s)$, $\hat{u}_i = h_i(c'^n_1, \hat{u}^{i-1}_1)$. In this case, the hard decision function $h_i$ is defined as: (i) $\forall \frac{W^{(i)}_n(c'^n_1, \hat{u}^{i-1}_1 | u_i = 0)}{W^{(i)}_n(c'^n_1, \hat{u}^{i-1}_1 | u_i = 1)} \geq 1$, $h_i(c'^n_1, \hat{u}^{i-1}_1) = 0$; (ii) otherwise, $h_i(c'^n_1, \hat{u}^{i-1}_1) = 1$. In other words, if the index $i$ of $W^{(i)}_n$ is not in the secret information set $\mathcal{A}(s)$, then the decoder knows that $\hat{u}_i = u_i = 0$.

3) After the SC decoder maps $\boldsymbol{c}'$ into $\hat{\boldsymbol{u}} = \hat{u}^n_1$, Bob can obtain the message as $\boldsymbol{m} = \boldsymbol{u}_{\mathcal{A}(s)}S$.

As shown in the above steps, $\mathcal{A}(s)$ is needed to execute the SC decoding. Therefore, it is impossible for any adversary to correct the intentional errors without knowing $\mathcal{A}(s)$. Fig. 2 shows the block diagram of the PKC-PC.

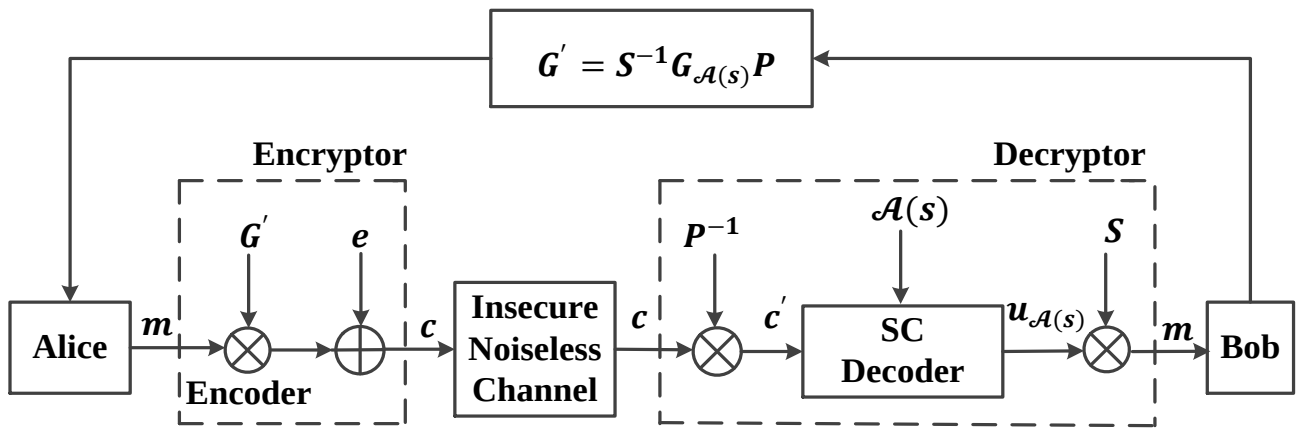

Fig. 2. Flowchart of the PKC-PC.

## IV. Efficiency Assessment

In this section, we measured the error performance, the key size and the computational complexity to evaluate the PKC-PC's efficiency level.

### A. Error Performance

As mentioned in Sec. II., the SC decoder estimates the value of $i$-th input bit, denoted by $\hat{u}_i$, given the received vector $\boldsymbol{y} = y^n_1$ and the prior evaluated input bits $\hat{u}_1, \hat{u}_2, \cdots, \hat{u}_{i-1}$ [27]. Therefore, to investigate the Hamming weight's upper bound of $\boldsymbol{e}$, the worst case in terms of error correction capability is considered for the SC decoding, i.e., when the erasure burst has occurred.

*Theorem [41]:* Consider a polar code of length $n = 2^m$, which is constructed for transmission over a BSM channel $W$. In this case, if an erasure burst of length at least $2\sqrt{n} - 1$ occurs, the SC decoder always fails to obtain the estimated message with probability of at least 0.5.

*Proof.* In [41].

The aforementioned theorem makes an upper bound on the error correction capability. For instance, in the PKC-PC, using polar code of length $n = 1024$ under SC decoding, the Hamming weight's upper bound of $\boldsymbol{e}$ is equal to 63.

### B. Key Size

In this section, we measure the public and private key lengths of the PKC-PC by using $(1024, 768)$ polar code as follows:

1) In the CCA2-secure variants, the encryption matrix can be considered in the systematic form which occupies $k(n-k)$ bits instead of $kn$ bits. As the PKC-PC is CCA2-secure (see Sec. VI.D), we can exploit the systematic encryption matrix $\mathcal{K}_{pub} = S^{-1}G_{\mathcal{A}(s)}P = [I_k|Q]$ as the public key. In this case, the public key length $\mathcal{M}_{pub}$ is approximately equal to 24.58 kbytes.

2) The PKC-PC's private key includes a set $\{\mathcal{A}^c(s), P\}$, in which $\mathcal{A}^c(s)$ is stored instead of $G_{\mathcal{A}(s)}$ and $S$. The maximum sub-channel index, i.e., $n = 1024$, may be in $\mathcal{A}^c(s)$ which requires 10 bits to store in binary. Therefore, the required memory to save $\mathcal{A}^c(s)$ is computed as $\mathcal{M}_{\mathcal{A}^c(s)} \leq 10(n-k) = 2.56$ kbits. The required memory to save $P_{n\times n}$ is obtained as $\mathcal{M}_P = n(n-k) = 32.77$ kbits. Hence, the private key length's upper bound is computed as $\mathcal{M}_{pri} = \mathcal{M}_{\mathcal{A}^c(s)} + \mathcal{M}_P \leq 32.77$ kbytes.

Table I compares the number of equivalent codes ($\mathcal{N}_C$), $\mathcal{M}_{pub}$, $R$ and the upper bound on $W_H(\boldsymbol{e})$ of PKC-PC with the McEliece scheme. It is obvious that although PKC-PC has larger $R$ and $k$, but its $\mathcal{M}_{pub}$ is smaller in comparison to the McEliece cryptosystem. Moreover, because of randomly choosing $k$ sub-channels among $n$ good sub-channels, the equivalent polar codes' number with the proposed parameters is computed as $\mathcal{N}_C \approx 2^{826}$ (see Sec. VI.A), which is much larger than the equivalent Goppa codes' number in the McEliece scheme.

TABLE I

COMPARING THE EFFICIENCY OF THE PKC-PC AND MCELIECE SCHEME.

| Scheme | McEliece [5] | PKC-PC | | |
|---|---|---|---|---|
| Code | Goppa | Polar | | |
| $(n, k)$ | $(1024, 524)$ | $(256, 192)$ | $(1024, 768)$ | $(1024, 921)$ |
| $\mathcal{N}_C$ | $2^{498}$ | $\approx 2^{204}$ | $\approx 2^{826}$ | $\approx 2^{478}$ |
| $\mathcal{M}_{pub}$ | 65.5 kbytes | 1.5 kbytes | 24 kbytes | 11.58 kbytes |
| $R$ | 0.512 | 0.75 | 0.75 | 0.9 |
| Upper bound on $W_H(e)$ | Patterson decoding | SC decoding | SC decoding | SC decoding |
| | 50 | 31 | 63 | 63 |
| Security Level | $2^{64.2}$ | $2^{79.96}$ | $2^{140.63}$ | $2^{247.98}$ |

### C. Computational Complexity

The computational complexity of the PKC-PC includes two parts: (i) encryption complexity ($\mathrm{C}_{Enc}$); and (ii) decryption complexity ($\mathrm{C}_{Dec}$). Encryption is performed by computing the product $\boldsymbol{m}G'$ and then adding the intentional error vector $\boldsymbol{e}$. Therefore, the encryption complexity can be expressed as $\mathrm{C}_{Enc} = \mathrm{C}_{mul}(\boldsymbol{m}G') + \mathrm{C}_{add}(\boldsymbol{e})$, where $\mathrm{C}_{mul}(\boldsymbol{m}G') = \mathcal{O}(k(n-k))$ is the complexity of multiplying $\boldsymbol{m}$ by the systematic encryption matrix $G' = [I_k|Q]$. Note that by using CCA2-secure conversion, the encryption matrix can be put in systematic form. In this case, $\mathrm{C}_{mul}(\boldsymbol{m}G')$ is reduced from $\mathcal{O}(kn)$ to $\mathcal{O}(k(n-k))$. Moreover, $\mathrm{C}_{add}(\boldsymbol{e}) = \mathcal{O}(n)$ is the needed binary operations' number for addition of $n$-bit $\boldsymbol{e}$. Although for a CCA2-secure variant implementation, the complexity of applying some proper scrambling operations on $\boldsymbol{m}$ before multiplication by $G'$ should be computed. The decryption complexity of PKC-PC is computed as $C_{Dec} = C_{mul}(\boldsymbol{c}P^{-1}) + C_{SC}(\boldsymbol{c}') + C_{mul}(\boldsymbol{u}_{\mathcal{A}(s)}S)$, where $C_{mul}(\boldsymbol{c}P^{-1}) = \mathcal{O}(n)$ is the needed binary operations' number to perform the multiplication of $n$-bit ciphertext $\boldsymbol{c}$ by the inverse of $P$. Also, the SC decoding's complexity is computed as $C_{SC}(\boldsymbol{c}') = \mathcal{O}(nlogn)$ [27]. Furthermore, the number of required binary operations for multiplying the $k$-bit vector $\boldsymbol{u}_{\mathcal{A}(s)} = \boldsymbol{m}S^{-1}$ by $S$ is computed as $C_{mul}(\boldsymbol{u}_{\mathcal{A}(s)}S) = \mathcal{O}(k^2)$.

## V. Formal Security Assessment

The security assessment of the PKC-PC is divided into two sections: (a) security reduction; (b) practical security. In this section, by using the security reduction proposed in [3, 42] for the original McEliece cryptosystem based on Goppa codes, we provide the reduction regarding the PKC-PC. We demonstrate the NP-completeness of some new variants of the hard decoding problem which are fitted to the specific polar codes' parameters. In addition, we provide a reduction proof regarding the PKC-PC. It implies that an attacker that is able to attack the PKC-PC is able to solve the new variants of hard decoding problem with a similar effort. Consider $\mathcal{C}$ as a binary polar code with length $n = 2^m$. Consider $t$ as an error correcting capability of $\mathcal{C}$ and $\omega$ as a positive integer whose magnitude is less than $t$. In the presented system, the adversary is encountered to specify $\boldsymbol{e}$ given a vector $\boldsymbol{c} = \boldsymbol{m}G' + \boldsymbol{e}$. Since the Hamming weight $\omega$ of intentional error vector $\boldsymbol{e}$ is less than $t$, the attacker performs a low weight word search algorithm to detect $\boldsymbol{e}$. In the sequel, we have shown that no proper algorithm exists to obtain $\boldsymbol{e}$ by the adversary. As a matter of fact, PKC-PC's security is reduced to solve the NP-complete problems, called polar parameterized syndrome decoding (PPSD) and polar parameterized codeword existence (PPCE). In fact, an NP-complete problem, called three-dimensional matching (TDM), is reduced to each of them. The PPSD and PPCE problems should be fixed to the $(n, k)$ polar codes' properties, i.e., $n = 2^m$, $k = n/R$ and $t = 2\sqrt{n} - 1$. It is enough to prove that none of PPSD and PPCE can be solved efficiently to ensure that no efficient attacker exists against the PKC-PC. Let $\mathcal{P}_{n,k}$ be the $(n, k)$ polar code family whose $k$ rows in their generator matrices are selected from the $n$ rows of $G_n$. Also, assuming that $\mathcal{H}_{n,r}$ is the set of all $n \times r$ matrices whose $r = n - k$ columns are selected from the columns of $G'_n = \begin{bmatrix} G_{\mathcal{A}(s)} \\ G_{\mathcal{A}^c(s)} \end{bmatrix}$.

*Problem 1 [3]. Three-Dimensional Matching (TDM)*
*Instance:* a subset $U \subseteq T \times T \times T$, where $T$ is a finite set.
*Question:* is there a set $W \subseteq U$ such that $|W| = |T|$, and no two elements of $W$ agree in any coordinate?

*Problem 2 [29]. Polar Parameterized Syndrome Decoding (PPSD)*
*Instance:* the parameters $\mathcal{H}_{n,r}$, $r = n - k$, a matrix $H \in \mathcal{H}_{n,r}$, a vector $\boldsymbol{s} \in F_2^r$ and a nonnegetive integer $\omega = 2\sqrt{n} - 1$.
*Question:* find $\boldsymbol{y} \in F_2^n$ with $w_H(\boldsymbol{y}) = 2\sqrt{n} - 1$ in such a way that $\boldsymbol{y}H = \boldsymbol{s}$?

*Proposition 1.* The PPSD problem is NP-complete.

*Proof.* Inspired by the presented approaches in [3, 42] and by reducing the TDM problem to PPSD problem, it can be demonstrated that PPSD problem is NP-complete. Consider $B$ as an $|U| \times |3T|$ incidence matrix. Each row of $B$ includes three 1s, one for each of the coordinate in the corresponding triple. Therefore, detecting an effective solution for the TDM problem relates to being a set of $|T|$ rows whose addition in $GF(2)$ yields an all one vector. As illustrated in Fig. 3, the matrix $B$ is expanded to $H_1$ of size $n \times r$. To perform such expansion, $n' = n - |U|$ full-zero rows and $r' = r - 3|T|$ full-zero columns is added to $B$. Such extension is performed to put the matrix $B$ suitable for the properties of PPSD problem.

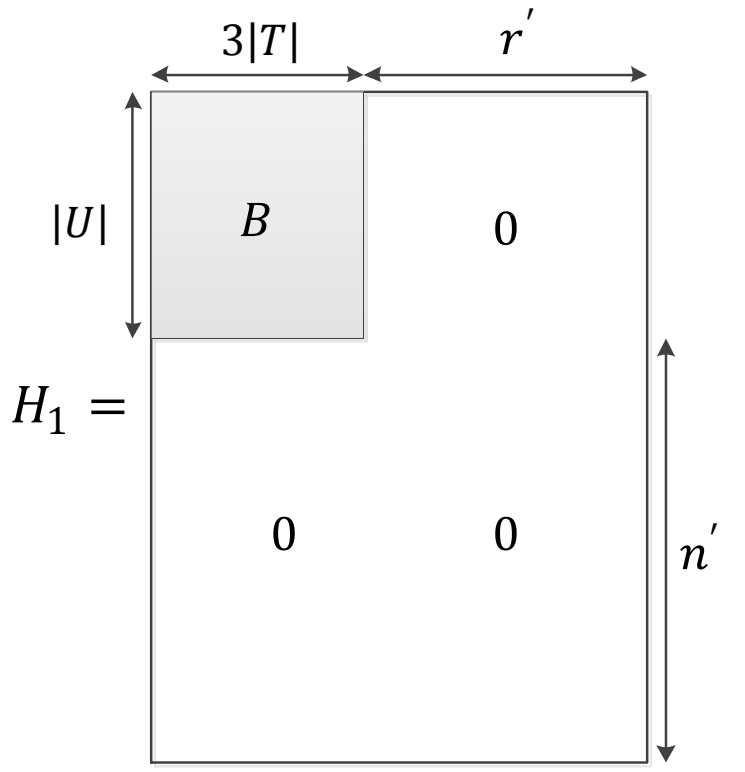

Fig. 3. Matrix $H_1$ [42] used to reduce TDM problem to PPSD problem.

Now, assuming that a polynomial time algorithm exists which can solve any sample of PPSD problem. The matrix $H_1$ and the syndrome $\boldsymbol{s} = (1, \cdots, 1, 0, \cdots, 0)$ consisting of $3|T|$ ones followed by $r'$ zeros are the inputs of this algorithm. By executing this algorithm, we can realize in polynomial time whether the $|T|$ triple set in $B$ is a matching. When the $|T| = 2\sqrt{n} - 1$ rows, the sum is the full-one vector is chosen from the $|U|$ top rows of $H_1$. Solving a PPSD problem in a polynomial-time gives a polynomial solution for TDM problem. This implies a polynomial-time solution for every NP problem which in turn demonstrates that PPSD problem is NP-complete. ∎

*Problem 3 [29]. Polar Parameterized Codeword Existence (PPCE)*
*Instance:* a binary matrix $H_{n \times r}, n = 2^m$, $r = n - k$ and a positive integer $\omega = 2\sqrt{n} - 1$.
*Question:* is there a codeword $\boldsymbol{x}$ of Hamming weight at most $\omega = 2\sqrt{n} - 1$ such that $\boldsymbol{x}H = 0$?

*Proposition 2.* the PPCE problem is NP-complete.

*Proof.* To prove the NP-completeness of PPCE problem, first the matrix $C$ (Fig. 4-a) is constructed [42]. Then, by inserting $n'' = n - 3|T|(|U| + 1) - |U|$ full zero rows and $r'' = r - 3|T|(|U| + 1)$ full zero columns to matrix $C$, it is expanded to matrix $H_2$ of size $n \times r$ (Fig. 4-b). This extension is performed to get $B$ fitted to the PPCE problem's properties. As a matter of fact, $C$ is a $(3|T||U| + 3|T| + |U|) \times (3|T||U| + 3|T|)$ matrix in which $|U|$ first rows include the matrix $B$ followed by $3|T|$ copies of the identity matrix $I_{|U|}$. Moreover, the $3|T|(|U| + 1)$ last rows of $C$ include an identity matrix $I_{3|T|(|U|+1)}$. Imagine that a polynomial-time algorithm exists which can solve any PPCE problem's instance. Now, $H_2$ and $\omega = 2\sqrt{n} - 1$ are considered as the PPCE problem's input.

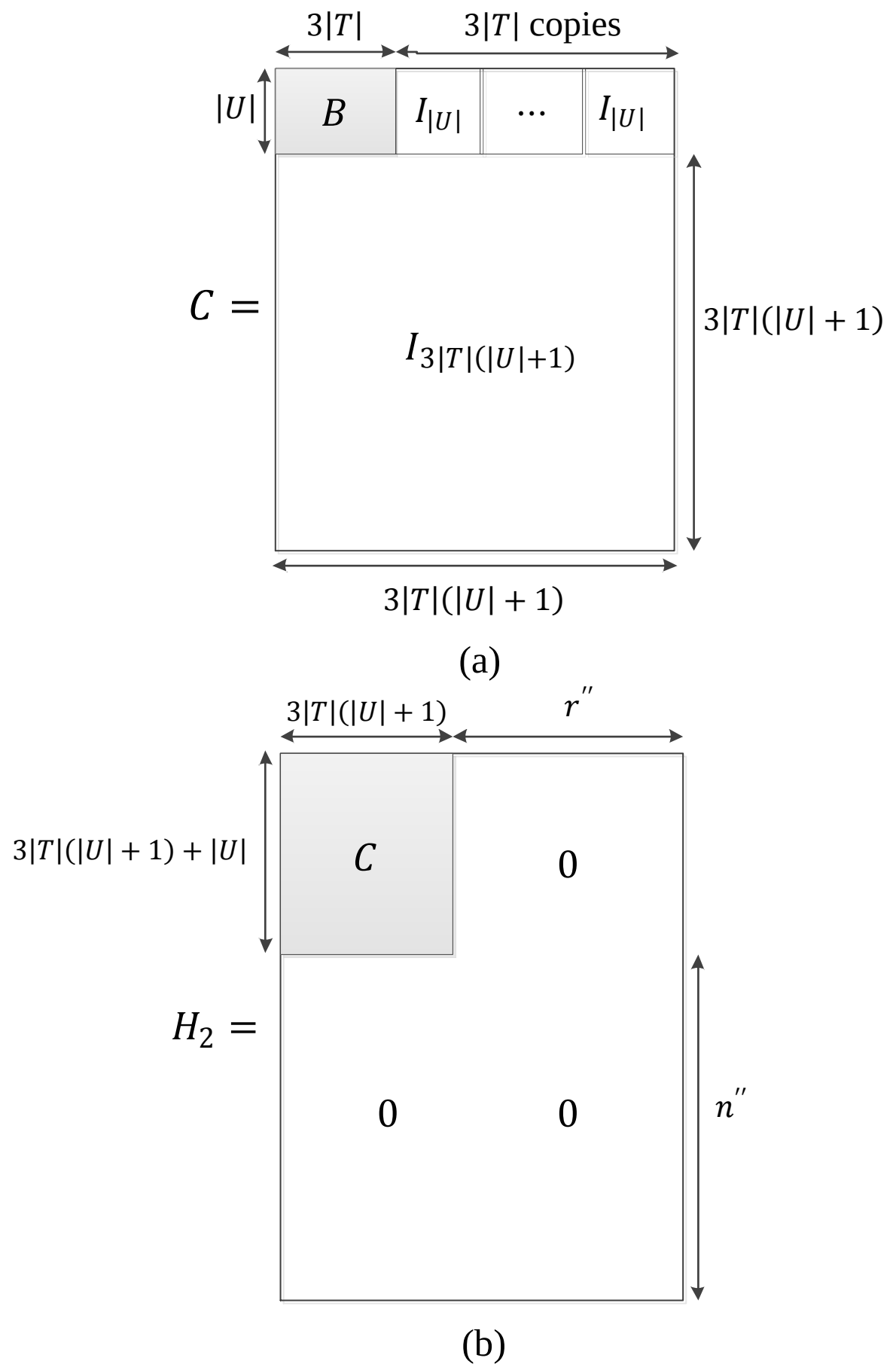

Fig. 4. (a) Matrix $C$ [42] used to reduce TDM problem to Subspace Weights problem. (b). Matrix $H_2$ is applied to reduce TDM problem to PPCE problem.

There is need to realize the word with Hamming weight $3|T|^2 + 4|T| = 2\sqrt{n} - 1$. In this way, the height $n = (3|T|^2 + 4|T| + 1)^2/4$ of $H_2$ is polynomial in $|T|$ and the extension from $C$ to $H_2$ is feasible. Moreover, suppose that $\boldsymbol{y} = (y_1, y_2, \cdots, y_{3|T|(|U|+1)+|U|}, 0, \cdots, 0)$ where $n''$ rightmost coordinates are zeros in such a way that $\boldsymbol{y}H_2 = 0$. Consider $\boldsymbol{y}_0 = (y_1, y_2, \cdots, y_{|U|})$ and $\boldsymbol{y}_1 = (y_{|U|+1}, y_{|U|+2}, \cdots, y_{3|T|(|U|+1)+|U|})$ as the $\boldsymbol{y}$'s subvectors. According to Fig. 4 (a, b), it is evident that $w_H(\boldsymbol{y}_1) = w_H(\boldsymbol{y}_0 A) + 3|T| w_H(\boldsymbol{y}_0)$. By adding $w_H(\boldsymbol{y}_0)$ to both sides of this equation, we have $w_H(\boldsymbol{y}) = w_H(\boldsymbol{y}_0 A) + (3|T| + 1) w_H(\boldsymbol{y}_0)$. In fact, $0 \le w_H(\boldsymbol{y}_0 A) \le 3|T|$ and $w_H(\boldsymbol{y}_0)$ can be specified from $w_H(\boldsymbol{y})$. When $w_H(\boldsymbol{y})$ is divided by $3|T| + 1$, $w_H(\boldsymbol{y}_0 A)$ and $w_H(\boldsymbol{y}_0)$ are the remainder and quotient, respectively. If $w_H(\boldsymbol{y}) = 3|T|^2 + 4|T|$, we have $w_H(\boldsymbol{y}_0 A) = 3|T|$ and $w_H(\boldsymbol{y}_0) = |T|$. Hence the code with parity check matrix $H_2$ has a word of Hamming weight $3|T|^2 + 4|T|$ if and only if the set of $|T|$ triples in $B$ has a matching [3]. In fact, a solution to the PPCE problem is a sum of $4|T|^2 + 3|T|$ rows summing to 0. It is a solution to TDM problem and demonstrates that the PPCE problem is NP-complete. ∎

*Proposition 3.* Breaking the Polar variant of the McEliece cryptosystem is not easier than solving the decoding problem for a random code.

*Proof.* This concludes from Propositions 1 and 2. ∎

## VI. Practical Security Assessment

In this section, we investigate the practical attacks against the PKC-PC. Generally, two types of practical attacks can be considered for the PKC-PC [43, 44]: (i) Structural attacks (key recovery attacks) whose aim is either at recovering the secret generator matrix $G_{\mathcal{A}(s)}$ of the employed polar code from the public key $G' = S^{-1} G_{\mathcal{A}(s)} P$ or also distinguishing the public key $G'$ from a random matrix (which invalidates the reduction proof); (ii) Decoding (message recovery) attacks whose aim is to decode a noisy codeword that contains a message $\boldsymbol{m}$ without exploiting any obvious structure of the secret generator matrix $G_{\mathcal{A}(s)}$.

### A. *Brute Force Attack*

Brute force attack is a kind of structural attack in which, all probable keys are searched and investigated consistently until the proper key is detected. However, this attack is doomed to fail if the space of private key set is large enough. Therefore, the secret code employed in the PKC-PC should be randomly chosen among a very large class of equivalent polar codes. The original McEliece cryptosystem using Goppa code, is immune against this attack. In the PKC-PC, due to random selection of $k$ sub-channels from $n$ good sub-channels, the equivalent $(n, k)$ polar codes' number and its dual are obtained as $\mathcal{N}_{\mathcal{C}} = \binom{n}{k}$ and $\mathcal{N}_{\mathcal{C}^\perp} = \binom{n}{n-k}$, respectively. This approach produces very large set of equivalent polar codes. For example, given a $(256, 192)$ polar code, $\mathcal{N}_{\mathcal{C}}$ is approximately equal to $2^{204}$. In addition, there are so many possibilities for the nonsingular and permutation matrices used in the PKC-PC. The number of binary nonsingular scrambler matrices is equal to the number of all possible submatrices $G_{i,j}$ of $G_n$ with indices $i, j \in \mathcal{A}(s)$, this means that $\mathcal{N}_S = \mathcal{N}_{\mathcal{C}} = \binom{n}{k}$. If $n$ and $k$ are properly chosen, $\mathcal{N}_{\mathcal{C}}$ is large enough. In this case, an adversary cannot detect $G_{\mathcal{A}(s)}$ in polynomial-time. The number of binary permutation matrices $P_{n \times n}$ is computed as $\mathcal{N}_P = \mathcal{N}_{P'} . \mathcal{N}_{P''} = \binom{n}{k} \times (n - k)$. Table II shows the average number of equivalent polar codes, nonsingular and permutation matrices for various code lengths $n$, dimension $k$ and rates $R = 0.75$ and $R = 0.9$. As shown in this table, due to the large parameters used in the PKC-PC, it is impossible to find $S$, $P$ and $G_{\mathcal{A}(s)}$ in polynomial time.

Table II

The Average Number of Equivalent Polar Codes, Nonsingular and Permutation Matrices for Various Code Lengths, Dimensions and Rates 0.75 and 0.9.

| $(n, k)$ | $R$ | $\mathcal{N}_{\mathcal{C}}$ | $\mathcal{N}_S$ | $\mathcal{N}_P$ |
|---|---|---|---|---|
| (256,192) | 0.75 | $2^{204}$ | $2^{204}$ | $2^{509}$ |
| (256,230) | 0.9 | $2^{118}$ | $2^{118}$ | $2^{206}$ |
| (512,384) | 0.75 | $2^{410}$ | $2^{410}$ | $2^{538}$ |
| (1024,768) | 0.75 | $2^{825}$ | $2^{825}$ | $2^{1081}$ |
| (2048,1536) | 0.75 | $2^{1656}$ | $2^{1656}$ | $2^{2168}$ |
| (4096,3072) | 0.75 | $2^{3317}$ | $2^{3317}$ | $\gg 2^{80}$ |

### B. Key Recovery Attack

In the distinguishing attack as a kind of algebraic attack, there is need to recognize the public key matrix from a randomly binary matrix by applying a distinguisher. This distinguisher, in its naive form can only invalidate the security reductions, and it can be more powerful if the distinguisher can reveal the hidden structure of the secret code. In [45], a deterministic distinguisher is proposed which is allowed to distinguish the matrix of a Goppa code from a random matrix. In fact, such distinguisher can solve Goppa code distinguishing (GCD) problem in polynomial-time for high code rates (near 1). The key ingredient of this method is an algebraic characterization of the key recovery problem and its idea is to consider the dimension of the solution space of a linearized system resulting from a particular polynomial system. We recall that the existence of such a distinguisher does not undermine the security of original McEliece cryptosystem. It is demonstrated that their security could not be reduced to the difficulty of random decoding of a linear code by means of GCD assumption. This kind of attacks, have better performance on some other cryptosystems using non binary Goppa codes [23, 24] and also generalized Reed Solomon (GRS) codes [44] since it leads to recovery of the secret codes. However, due to the following reasons, these distinguishing attacks are ineffective against the PKC-PC: (i) it is unable to recognize the public key matrix of the PKC-PC from a randomly generated one, i.e., public key is resistant to this attack, because the public key $G'$ is not the generator matrix of polar codes. This is because of multiplying $S^{-1}$ to $G_{\mathcal{A}(s)}P$; (ii) the recognizer cannot work on subspaces of the code, hence it is impossible to detect the subspace that the attacker needs.

In [35], Bardet et al. present a new key recovery attack by which Shrestha-Kim [34] polar code-based public key cryptosystem is broken. In fact, a new family of codes, called decreasing monomial codes, is suggested which consists as a special case, Reed-Muller codes and Polar codes. By means of these codes, low weight codewords in underlying polar code and its dual are obtained. Moreover, it is possible to recover the permuted polar code by enhancing all the information required for decrypting any message. It is shown that the code equivalence problem for binary polar codes can be solved efficiently by a more complicated algorithm with the help of the following four steps: The first step is searching for minimum weight codewords using Stern [46] and Dumer [47] algorithms. The second step is shortening the code with respect to the low weight codewords found in the first step and in the dual code. In the third step, by characterizing the permutation group of polar codes together with the low-weight codewords found in Step 2, it is possible to find, among the codewords found in Step 1, a subset of codewords which up to equivalence by the permutation group. The fourth step is to puncture the code with respect to the support of an element of minimum weight in this last subset of codewords gives a code of small length whose structure is known up to code equivalence. The code equivalence problem is then solved in this case and is used to recover step by step the underlying polar codes.

It is shown that the only way to avoid this key recovery attack is to look for polar code parameters for which finding minimum weight codewords is unable either in the code or in its dual. This would require changing significantly the parameters proposed in Shrestha- Kim scheme that would make such scheme much less attractive. However, this attack is not applicable to the PKC-PC since we select a special kind of random subcode of polar codes instead of naïve form. This proper selection doesn't allow solving of code equivalent problem. In fact, since the number of code equivalent for used polar code $\mathcal{C}$ and its dual, i.e., $\mathcal{N}_{\mathcal{C}}$ and $\mathcal{N}_{\mathcal{C}^{\perp}}$, are large enough, the PKC-PC is immune against such key recovery attack.

### C. Information Set Decoding Attack

Information set decoding (ISD) attack is the most powerful kind of decoding attack that usually determines the work factor of code-based cryptosystems. ISD attack attempts to find the error vector $\boldsymbol{e}$ in ciphertext by searching for the codewords with minimum weight in the given code extended by the received codeword, that is, the code described by the generator matrix $\begin{bmatrix} G \\ c \end{bmatrix}$. This approach uses an ISD algorithm to search for the minimum weight codeword which is equivalent to find $\boldsymbol{e}$. A naive form of ISD attack was introduced by Prange [48] and used in the original McEliece cryptosystem [5]. From then on, many subsequent variants were introduced [49-53]. One important step in the development of ISD attack is Stern attack [46] in which a probabilistic and explicit algorithm is presented to find low-weight codeword in an $(n,k)$ binary linear code. In this paper, we consider the Stern attack [46] to analyze the strength of the PKC-PC against ISD attack.

The inputs of this attack are as follows: (i) an integer $\omega \geq 0$; (ii) an $(n-k)\times n$ parity check matrix $H$ or a $k\times n$ generator matrix $G$ of an $(n,k)$ polar code. Let us denote the work factor of ISD attack in a $(n,k)$ binary linear code to find a single codeword of weight $\omega$ by $\mathrm{WF_{isd}}(n,k,\omega)$. By applying the Stern algorithm, the ISD attack's work factor is obtained as $\mathrm{WF_{isd}}(n,k,\omega) = Cost_{ST}/P_{ST}$, where the number of binary operations required to perform each iteration of algorithm is calculated as $Cost_{ST} = \frac{1}{2}(n-k)^2(n+k) + 2\binom{k/2}{p}p\ell + 2p(n-k)\binom{k/2}{p}^2/2^{\ell}$ and the success probability of finding a single codeword of weight $\omega$ is $\mathrm{P}_{ST} = \binom{k/2}{p}^2\binom{n-k-\ell}{\omega-2p}\binom{n}{\omega}^{-1}$, $0 \leq p \leq \omega$ and $0 \leq \ell \leq n-k$ are two integers as the algorithm parameters whose size is determined in such a way that the complexity of attack is minimized [46].

TABLE III
WORK FACTOR ($\log_2$) OF ISD ATTACKS ON POLAR CODES WITH VARIOUS CODE LENGTHS AND DIMENSIONS FOR $R = 0.75$.

| $(n,k)$ | $(p,\ell)$ | $w_H(\boldsymbol{e})$ | WF($log_2$) | PK (kByte) |
|---|---|---|---|---|
| (256,192) | (2,8) | 31 | 79.96 | 1.5 |
| (512,384) | (3,22) | 44 | 104.61 | 6 |
| (1024,768) | (5,39) | 63 | 140.63 | 24 |
| (2048,1536) | (7,59) | 89 | 190.19 | 96 |
| (4096,3072) | (15,124) | 127 | 266.34 | 384 |

TABLE IV
WORK FACTOR ($\log_2$) OF ISD ATTACKS ON POLAR CODES WITH VARIOUS CODE RATES AND DIMENSIONS FOR $n = 1024$, $w_H(\boldsymbol{e}) = 63$.

| $R$ | $k$ | $(p,\ell)$ | WF($log_2$) | $\mathcal{N}_C(log_2)$ | PK (kByte) |
|---|---|---|---|---|---|
| 0.5 | 512 | (3,27) | 74.90 | 1018,67 | 32 |
| 0.6 | 614 | (3,27) | 94.82 | 989.19 | 30.73 |
| 0.7 | 717 | (3,27) | 122.41 | 897.00 | 26.87 |
| 0.75 | 768 | (5,39) | 140.63 | 825.63 | 24 |
| 0.8 | 819 | (9,61) | 163.70 | 734.65 | 20.49 |
| 0.9 | 921 | (5, 1) | 247.98 | 477.56 | 11.58 |

Some sets of polar codes parameters of $R = 0.75$ with corresponding security level which are calculated by Stern algorithm is shown in Table III. Moreover, some other sets of polar codes parameters of length $n = 1024$ for various code rates are given in Table IV. The results of Table IV show that the polar codes have a wide range of flexibility in code rate which make it possible to decrease the public key size at the cost of decreasing $\mathcal{N}_C$. Since $\mathcal{N}_C$ is still below the complexity of ISD attack, the work factor is determined by ISD attack.

### D. CCA2-Secure Version of the PKC-PC

As mentioned earlier, in the PKC-PC, we use a systematic encryption matrix $G' = [I_k|Q]$ as the public key which have the following advantages [43]: (i) the size of public key becomes much smaller, i.e., it requires $k(n-k)$ bits instead of $kn$ bits; (ii) the encryption is faster because it suffices to multiply the plaintext $\boldsymbol{m}$ by $k \times (n-k)$ submatrix $Q$ instead of $k \times n$ encryption matrix $G'$; (iii) the decryption is faster, because the message is a prefix of the ciphertext and can be recovered easily. However, using a systematic encryption matrix can lead to the loss of security against adaptive chosen ciphertext attack (CCA2). In such attacks, given a ciphertext $\boldsymbol{c}$ of a message $\boldsymbol{m}$, i.e., $\boldsymbol{c} = \boldsymbol{m}G' + \boldsymbol{e}$, the attacker inputs $\boldsymbol{c} + \boldsymbol{m}'G'$ to the decryption oracle for some $\boldsymbol{m}'$ and obtains the outputs of decryption oracle as $\bar{\boldsymbol{m}}$. Then, the attacker can recover the message as $\boldsymbol{m} = \bar{\boldsymbol{m}} - \boldsymbol{m}'$. Therefore, we should secure the PKC-PC against CCA2 to enable us use the systematic encryption matrix $G'$ without loss of security. In fact, the PKC-PC is CCA2-secure if an attacker with access to decryption oracle doesn't have any advantage in deciphering a given ciphertext $\boldsymbol{c}$. Also, indistinguishability against adaptive chosen ciphertext attacks (IND-CCA2) is achieved if Alice encrypts one of two messages $\boldsymbol{m}_1$ and $\boldsymbol{m}_2$, $\boldsymbol{m}_1 \neq \boldsymbol{m}_2$ to obtain a ciphertext $\boldsymbol{c}$ and has no advantage in distinguishing the message.

Several techniques were proposed to make the McEliece cryptosystem IND-CCA2 [54-58]. All suggested approaches in these conversions are based on scrambling the message inputs. In such way, any relation between two dependent messages which might be extracted by an attacker to recover the message is destroyed. It means that applying CCA2-secure conversion will enable us to perform a systematic generator matrix without loss of security. Two instances of generic conversions which can be applicable to the PKC-PC are Pointcheval conversion [55] and Fujisaki-Okamato conversion [54]. Although CCA2-secure scheme can be achieved by using the generic conversions [54, 55], they are not appropriate enough to be applied in the PKC-PC because these conversions add large amounts of redundancy to the ciphertexts. Instead, in the specific conversion, e.g., Kobara-Imai $\gamma$-conversion [56], a data redundancy is reduced even for the large parameters. Hence, we apply Kobara-Imai $\gamma$-conversion by which the data overhead is decreased compared with the generic conversions to have a CCA2-secure PKC-PC. It is indicated that breaking indistinguishability in the CCA2 model using Kobara-Imai $\gamma$-conversion is as difficult as breaking the McEliece scheme [56].

Another weakness of the PKC-PC is the malleability of the ciphertexts. In this case, the attacker can use the relation between two encrypted messages to determine the error bits. Let $\boldsymbol{m}_1$ and $\boldsymbol{m}_2$ be two messages that have a known relation $\Lambda$, e.g., $\Lambda(\boldsymbol{m}_1, \boldsymbol{m}_2) = \boldsymbol{m}_1 + \boldsymbol{m}_2$. Let $\boldsymbol{c}_1 = \boldsymbol{m}_1 G' + \boldsymbol{e}_1$ and $\boldsymbol{c}_2 = \boldsymbol{m}_2 G' + \boldsymbol{e}_2$ be the corresponding ciphertexts of $\boldsymbol{m}_1$ and $\boldsymbol{m}_2$, respectively. In this case, $\boldsymbol{c}_1 + \boldsymbol{c}_2 + \Lambda(\boldsymbol{m}_1, \boldsymbol{m}_2)$ has the Hamming weight of less than $2t = 4\sqrt{n} - 2$ and at least $k$ error-free positions of $\boldsymbol{m}_1 + \boldsymbol{m}_2$ can be revealed. This property allows an attacker to guess the error bits. A special case of related messages occurs in the message-resend attack in which the attacker can recover $\boldsymbol{e}_1 + \boldsymbol{e}_2 = \boldsymbol{c}_1 + \boldsymbol{c}_2$. Another attack is a reaction attack, a weaker version of CCA2, in which the attacker changes a few bits of ciphertext and watches the reaction of the legitimate receiver on these changed bits. If the receiver cannot decode the ciphertext and hence requests to resend it, the corresponding bits are not in error originally. This enables the attacker to obtain the error-free information in at most $k$ iterations. It should be noted that using the Kobara-Imai $\gamma$-conversion makes the PKC-PC secure against practical attacks such as, related message attack [59], message resend attack, reaction attack [60] and malleability attack.

## VII. CONCLUSION

This paper introduced a variant of the McEliece public key cryptosystem based on polar codes, called as PKC-PC. It has a number of benefits such as larger information rate and smaller public key length in comparison with the McEliece cryptosystem. By using Kobara-Imai's $\gamma$-conversion, we have attempted to have secure scheme against adaptive chosen ciphertext attacks. In this approach, we can convert the encryption matrix $G'$ to the systematic matrix which yields to reduce the public key length. We have shown that the PKC-PC's security is reduced to solve the NP-complete PPSD and PPCE problems. Also, the investigation's results show the flexibility of the PKC-PC. To design a secure and efficient PKC-PC, the parameters such as code length, code dimension and the Hamming weight of the error vector should be chosen in such a way that a suitable tradeoff will be performed between security and efficiency.